\documentclass[letterpaper]{article}
\usepackage{aaai20}
\usepackage{times}
\usepackage{helvet}
\usepackage{courier}
\usepackage{xcite}
\usepackage{amsmath,amssymb,amsfonts}
\usepackage{algorithmic}
\usepackage{graphicx}
\usepackage{multirow}
\usepackage{bigstrut}
\usepackage{enumitem}
\usepackage{subcaption}
\usepackage{mathtools}

\begin{document}
%
\title{Memory Augmented Graph Neural Networks for Sequential Recommendation}
\author{
Chen Ma\thanks{Work done as interns at Huawei Noah’s Ark Lab in Montreal.}\textsuperscript{\rm 1}{\normalfont,}
Liheng Ma\footnotemark[1]\textsuperscript{\rm 1,3}{\normalfont,}
Yingxue Zhang\textsuperscript{\rm 2}{\normalfont,}
Jianing Sun\textsuperscript{\rm 2}{\normalfont,}
Xue Liu\textsuperscript{\rm 1}
{\normalfont and} Mark Coates\textsuperscript{\rm 1} \\
\textsuperscript{\rm 1}McGill University, 
\textsuperscript{\rm 2}Huawei Noah's Ark Lab in Montreal,
\textsuperscript{\rm 3}Mila
\\
chen.ma2@mail.mcgill.ca, liheng.ma@mail.mcgill.ca, yingxue.zhang@huawei.com\\ jianing.sun@huawei.com,
xueliu@cs.mcgill.ca, mark.coates@mcgill.ca
}
\maketitle
\begin{abstract}
\begin{quote}
The chronological order of user-item interactions can reveal time-evolving and sequential user behaviors in many recommender systems. The items that users will interact with may depend on the items accessed in the past. However, the substantial increase of users and items makes sequential recommender systems still face non-trivial challenges: (1) the hardness of modeling the short-term user interests; (2) the difficulty of capturing the long-term user interests; (3) the effective modeling of item co-occurrence patterns. To tackle these challenges, we propose a memory augmented graph neural network (MA-GNN) to capture both the long- and short-term user interests. Specifically, we apply a graph neural network to model the item contextual information within a short-term period and utilize a shared memory network to capture the long-range dependencies between items. In addition to the modeling of user interests, we employ a bilinear function to capture the co-occurrence patterns of related items. We extensively evaluate our model on five real-world datasets, comparing with several state-of-the-art methods and using a variety of performance metrics. The experimental results demonstrate the effectiveness of our model for the task of Top-K sequential recommendation.
\end{quote}
\end{abstract}

\section{Introduction}
With the rapid growth of Internet services and mobile devices, personalized recommender systems play an increasingly important role in modern society. They can reduce information overload and help satisfy diverse service demands. Such systems bring significant benefits to at least two parties. They can: (i) help users easily discover products from millions of candidates, and (ii) create opportunities for product providers to increase revenue. 

On the Internet, users access online products or items in a chronological order. The  items a user will interact with in the future may depend strongly on the items he/she has accessed in the past. This property facilitates a practical application scenario---sequential recommendation. In the sequential recommendation task, in addition to the general user interest captured by all general recommendation models, we argue that there are three extra important factors to model: user \textit{short-term interests}, user \textit{long-term interests}, and \textit{item co-occurrence patterns}. The user short-term interest describes the user preference given several recently accessed items in a short-term period. The user long-term interest captures the long-range dependency between earlier accessed items and the items users will access in the future. The item co-occurrence pattern illustrates the joint occurrences of commonly related items, such as a mobile phone and a screen protector.

Although many existing methods have proposed effective models, we argue that they do not fully capture the aforementioned factors. First, methods like Caser~\cite{DBLP:conf/wsdm/TangW18}, MARank~\cite{DBLP:conf/aaai/YuZLZ19}, and Fossil~\cite{DBLP:conf/icdm/HeM16} only model the short-term user interest and ignore the long-term dependencies of items in the item sequence. The importance of capturing the long-range dependency has been confirmed by~\cite{DBLP:conf/kdd/BellettiCC19}.
Second, methods like SARSRec~\cite{DBLP:conf/icdm/KangM18} do not explicitly model the user short-term interest. Neglecting the user short-term interest prevents the recommender system from understanding the time-varying user intention over a short-term period. Third, methods like GC-SAN~\cite{DBLP:conf/ijcai/XuZLSXZFZ19} and GRU4Rec+~\cite{DBLP:conf/cikm/HidasiK18} do not explicitly capture the item co-occurrence patterns in the item sequences. Closely related item pairs often appear one after the other and a recommender system should take this into account.

To incorporate the factors mentioned above, we propose a memory augmented graph neural network (MA-GNN) to tackle the sequential recommendation task. This consists of a general interest module, a short-term interest module, a long-term interest module, and an item co-occurrence module.
In the general interest module, we adopt a matrix factorization term to model the general user interest without considering the item sequential dynamics.
In the short-term interest module, we aggregate the neighbors of items using a GNN to form the user intentions over a short period. These can capture the local contextual information and structure~\cite{DBLP:journals/corr/abs-1806-01261} within this short-term period. To model the long-term interest of users, we use a shared key-value memory network to generate the interest representations based on users' long-term item sequences. By doing this, other users with similar preferences will be taken into consideration when recommending an item. 
To combine the short-term and long-term interest, we introduce a gating mechanism in the GNN framework, which is similar to the long short-term memory (LSTM)~\cite{DBLP:journals/neco/HochreiterS97}. This controls how much the long-term or the short-term interest representation can contribute to the combined representation. In the item co-occurrence module, we apply a bilinear function to capture the closely related items that appear one after the other in the item sequence. 
We extensively evaluate our model on five real-world datasets, comparing it with many state-of-the-art methods using a variety of performance validation metrics. The experimental results not only demonstrate the improvements of our model over other baselines but also show the effectiveness of the proposed components.

To summarize, the major contributions of this paper are:
\begin{itemize}[leftmargin=*]
\item To model the short-term and long-term interests of users, we propose a memory augmented graph neural network to capture items' short-term contextual information and long-range dependencies.
\item To effectively fuse the short-term and long-term interests, we incorporate a gating mechanism within the GNN framework to adaptively combine these two kinds of hidden representations.
\item To explicitly model the item co-occurrence patterns, we use a bilinear function to capture the feature correlations between items.
\item Experiments on five real-world datasets show that the proposed MA-GNN model significantly outperforms the state-of-the-art methods for sequential recommendation.
\end{itemize}

\section{Related Work}

\subsection{General Recommendation}
Early recommendation studies largely focused on explicit feedback~\cite{DBLP:conf/kdd/Koren08}. The recent research focus is shifting towards implicit data~\cite{DBLP:conf/www/TranLLK19,DBLP:conf/kdd/LiS17}. Collaborative filtering (CF) with implicit feedback is usually treated as a Top-K item recommendation task, where the goal is to recommend a list of items to users that users may be interested in. It is more practical and challenging~\cite{DBLP:conf/icdm/PanZCLLSY08}, and accords more closely with the real-world recommendation scenario. Early works mostly rely on matrix factorization techniques~\cite{DBLP:conf/icdm/HuKV08,DBLP:conf/uai/RendleFGS09} to learn latent features of users and items. Due to their ability to learn salient representations, (deep) neural network-based methods~\cite{DBLP:conf/www/HeLZNHC17} are also adopted. Autoencoder-based methods~\cite{DBLP:conf/cikm/MaZWL18,DBLP:conf/wsdm/MaKWWL19} have also been proposed for Top-K recommendation. In~\cite{DBLP:conf/kdd/LianZZCXS18,DBLP:conf/ijcai/XueDZHC17}, deep learning techniques are used to boost the traditional matrix factorization and factorization machine methods.

\subsection{Sequential Recommendation}

The sequential recommendation task takes as input the chronological item sequence. A Markov chain~\cite{DBLP:conf/ijcai/ChengYLK13} is a classical option for modelling the data. For example, FPMC~\cite{DBLP:conf/www/RendleFS10} factorizes personalized Markov chains in order to capture long-term preferences and short-term transitions. Fossil~\cite{DBLP:conf/icdm/HeM16} combines similarity-based models with high-order Markov chains. TransRec~\cite{DBLP:conf/recsys/HeKM17} proposes a translation-based method for sequential recommendation. Recently, inspired by the advantages of sequence learning in natural language processing, researchers have proposed (deep) neural network based methods to learn the sequential dynamics. For instance, Caser~\cite{DBLP:conf/wsdm/TangW18} applies a convolutional neural network (CNN) to process the item embedding sequence. Recurrent neural network (RNN)-based methods, especially gated recurrent unit (GRU)-based methods~\cite{DBLP:conf/cikm/HidasiK18,DBLP:journals/corr/HidasiKBT15,DBLP:conf/cikm/LiRCRLM17} have been used to model the sequential patterns for the task of session-based recommendation. Self-attention~\cite{DBLP:conf/nips/VaswaniSPUJGKP17} exhibits promising performance in sequence learning and is starting to be used in sequential recommendation. SASRec~\cite{DBLP:conf/icdm/KangM18} leverages self-attention to adaptively take into account the interactions between items. Memory networks~\cite{DBLP:conf/wsdm/ChenXZT0QZ18,DBLP:conf/sigir/HuangZDWC18} are also adopted to memorize the items that will play a role in predicting future user actions.

However, our proposed model is different from previous models. We apply a graph neural network with external memories to capture the short-term item contextual information and long-term item dependencies. In addition, we also incorporate an item co-occurrence module to model the relationships between closely related items.

\section{Problem Formulation}
The recommendation task considered in this paper takes sequential implicit feedback as training data. The user preference is represented by a user-item sequence in chronological order, $ \mathcal{S}^{u}=(I_{1}, I_{2},...,I_{|\mathcal{S}^{u}|}) $, where $ I_{*} $ are item indexes that user $ u $ has interacted with. Given the earlier subsequence $ \mathcal{S}^{u}_{1:t} (t < |\mathcal{S}^{u}|) $ of $ M $ users, the problem is to recommend a list of $ K $ items from a total of $ N $ items ($ K < N $) to each user and evaluate whether the items in $ \mathcal{S}^{u}_{t+1:|\mathcal{S}^{u}|} $ appear in the recommended list.

\section{Methodology}

In this section, we introduce the proposed model, MA-GNN, which applies a memory augmented graph neural network for the sequential recommendation task. We introduce four factors that have an impact on the user preference and intention learning. Then we introduce the prediction and training procedure of the proposed model.

\begin{figure*}[ht]
    \centering
    \includegraphics[width=0.8\textwidth]{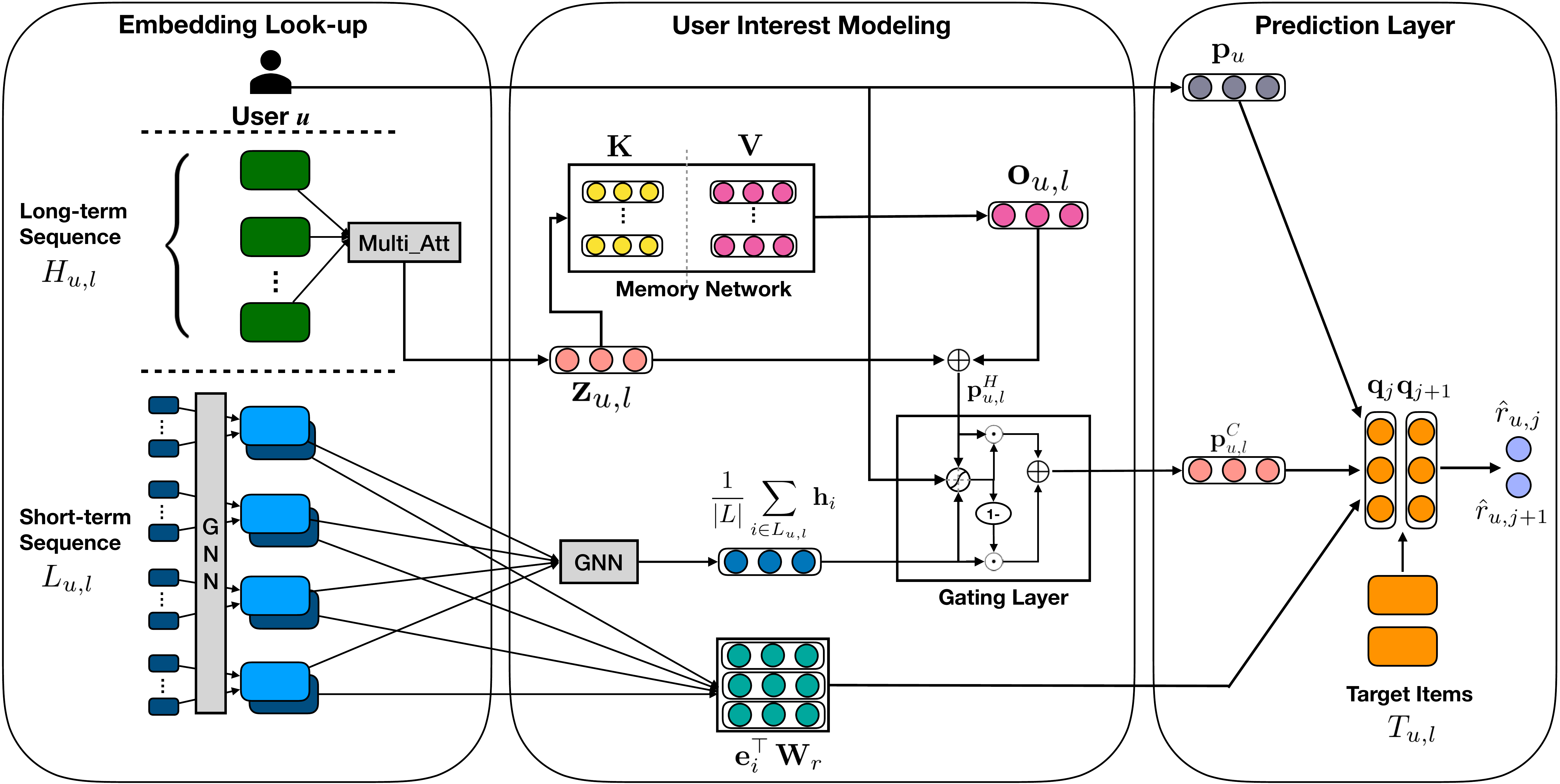}
    \caption{The model architecture of MA-GNN. $ \oplus $ denotes element-wise addition and $ \odot $ denotes element-wise multiplication.}
    \label{fig:model_figure}
\end{figure*}

\subsection{General Interest Modeling}
The general or static interest of a user captures the inherent preferences of the user and is assumed to be stable over time. To capture the general user interest, we employ a matrix factorization term without considering the sequential dynamics of items. This term takes the form
\[
\mathbf{p}_u^{\top} \cdot \mathbf{q}_j \,,
\]
where $ \mathbf{p}_u \in \mathbb{R}^{d} $ is the embedding of user $ u $, $ \mathbf{q}_j \in \mathbb{R}^{d} $ is the output embedding of item $ j $, and $ d $ is the dimension of the latent space.

\subsection{Short-term Interest Modeling}
A user's short-term interest describes the user's current preference and is based on several recently accessed items in a short-term period. The items a user will interact with in the near future are likely to be closely related to the items she just accessed, and this property of user behaviors has been confirmed in many previous works~\cite{DBLP:conf/wsdm/TangW18,DBLP:conf/cikm/HidasiK18,DBLP:conf/icdm/HeM16}. Therefore, it is very important in sequential recommendation to effectively model the user's short-term interest, as reflected by recently accessed items.

To explicitly model the user short-term interest, we conduct a sliding window strategy to split the item sequence into fine-grained sub-sequences. We can then focus on the recent sub-sequence to predict which items will appear next and ignore the irrelevant items that have less impact. For each user $ u $, we extract every $ |L| $ successive items as input and their next $ |T| $ items as the targets to be predicted, where $ L_{u,l}=(I_{l}, I_{l + 1},..., I_{l + |L| - 1}) $ is the $l$-th sub-sequence of user $ u $. Then the problem can be formulated as: in the user-item interaction sequence $ \mathcal{S}^{u} $, given a sequence of $ |L| $ successive items, how likely is it that the predicted items accord with the target $|T|$ items for that user. Due to their ability to perform neighborhood information aggregation and local structure learning~\cite{DBLP:journals/corr/abs-1806-01261}, graph neural networks (GNNs) are a good match for the task of aggregating the items in $ L_{u,l} $ to learn user short-term interests.

\begin{figure}[ht]
    \centering
    \includegraphics[width=0.5\textwidth]{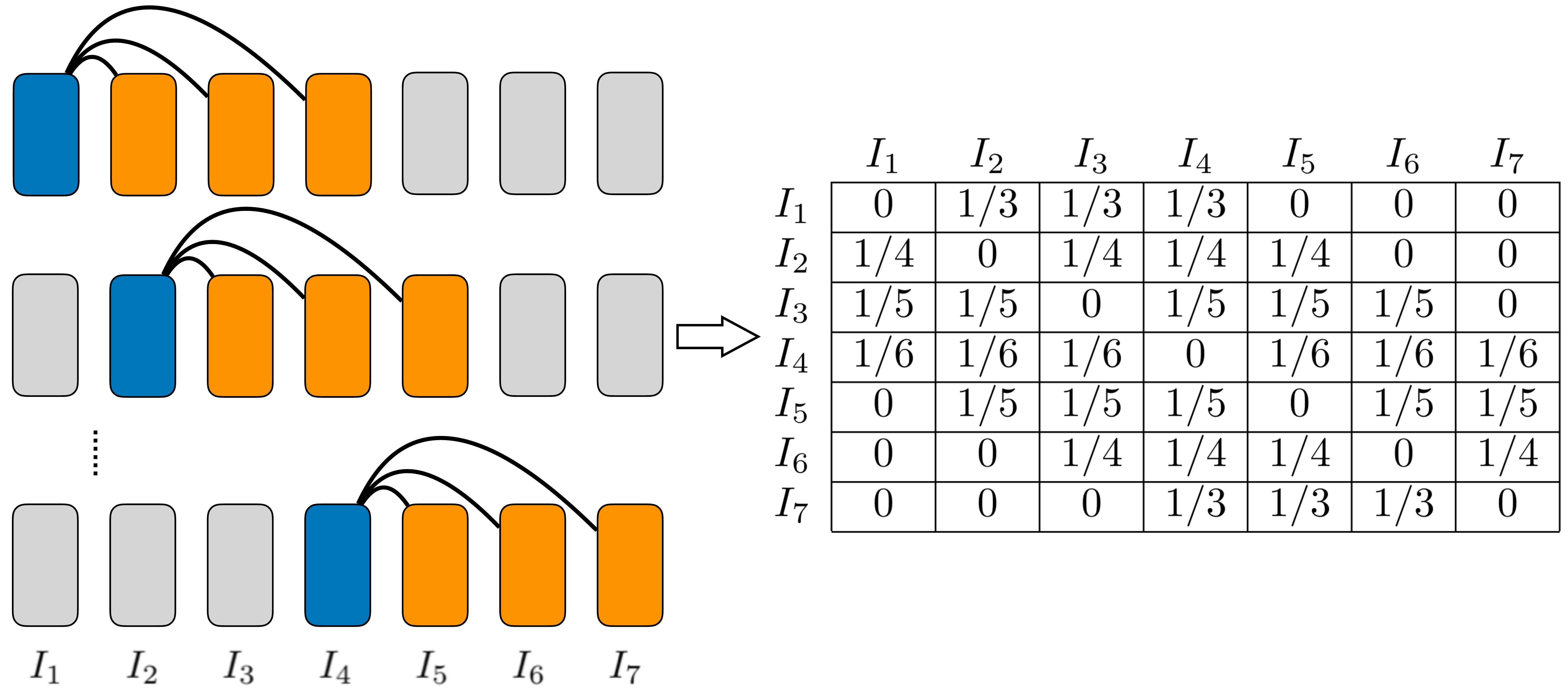}
    \caption{Item adjacent matrix construction example.}
    \label{fig:graph_construction}
\end{figure}

\textbf{Item Graph Construction}.
Since item sequences are not inherently graphs for GNN training, we need to build a graph to capture the connections between items. For each item in item sequences, we extract several subsequent items (three items in our experiments) and add edges between them. We perform this for each user and count the number of edges of extracted item pairs across all users. Then we row-normalize the adjacency matrix. As such, relevant items that appear closer to one another in the sequence can be extracted. An example of how to extract item neighbors and build the adjacency matrix is shown in Figure~\ref{fig:graph_construction}. We denote the extracted adjacency matrix as $ \mathbf{A} $, where $ A_{i,k} $ denotes the normalized node weight of item $ k $ regarding item $ i $. And the neighboring items of item $ i $ is denoted as $ \mathcal{N}_i $.

\textbf{Short-term Interest Aggregation}. To capture the user short-term interest, we use a two-layer GNN to aggregate the neighboring items in $ L_{u,l} $ for learning the user short-term interest representation. Formally, for an item $ j $ in the $l$-th short-term window $ L_{u,l} $, its input embedding is represented as $ \mathbf{e}_j \in \mathbb{R}^{d} $. The user short-term interest is then:
\begin{equation}
    \mathbf{h}_i = \tanh(\mathbf{W}^{(1)} \cdot [ \, \sum_{k \in \mathcal{N}_i} \mathbf{e}_k \, A_{i,k}; \mathbf{e}_i \, ]) \,, \forall i \in L_{u,l} \,,
\label{eq:GCN_aggregation1}
\end{equation}
\begin{equation}
    \mathbf{p}^{S}_{u, l} = \tanh(\mathbf{W}^{(2)} \cdot [ \, \frac{1}{|L|}\sum_{i \in L_{u,l}} \mathbf{h}_i; \mathbf{p}_{u} \, ]) \,,
\label{eq:GCN_aggregation2}
\end{equation}
where $ [\cdot \,; \,\cdot] \in \mathbb{R}^{2d} $ denotes vertical concatenation, $ \mathbf{W}^{(1)}, \mathbf{W}^{(2)} \in \mathbb{R}^{d \times 2d}$ are the learnable parameters in the graph neural network, and the superscript $ S $ denotes that the representation is from the user short-term interest. By aggregating neighbors of items in $ L_{u,l} $, $ \mathbf{p}^{S}_{u, l} $ represents a union-level summary~\cite{DBLP:conf/wsdm/TangW18,DBLP:conf/aaai/YuZLZ19} indicating which items are closely relevant to the items in $ L_{u,l} $. Based on the summarized user short-term interest, the items that a user will access next can be inferred.

However, directly applying the above GNN to make predictions clearly neglects the long-term user interest in the past $ H_{u,l}=(I_{1}, I_{2},...,I_{l-1}) $. There may be some items outside the short-term window $ L_{u,l} $ that can express the user preference or indicate the user state. These items can play an important role in predicting items that will be accessed in the near future. This long-term dependency has been confirmed in many previous works~\cite{DBLP:conf/kdd/LiuZMZ18,DBLP:conf/ijcai/XuZLSXZFZ19,DBLP:conf/kdd/BellettiCC19}. Thus, how to model the long-term dependency and balance it with the short-term context is a crucial question in sequential recommendation.

\subsection{Long-term Interest Modeling}
To capture the long-term user interest, we can use external memory units~\cite{DBLP:conf/nips/SukhbaatarSWF15,DBLP:conf/www/ZhangSKY17} to store the time-evolving user interests given the user accessed items in $ H_{u,l}=(I_{1}, I_{2},...,I_{l-1}) $. However, maintaining the memory unit for each user has a huge memory overhead to store the parameters. Meanwhile, the memory unit may capture information that is very similar to that represented by the user embedding $ \mathbf{p}_u $. Therefore, we propose to use a memory network to store the latent interest representation shared by all users, where each memory unit represents a certain type of latent user interest, such as the user interest regarding different categories of movies. Given the items accessed by a user in the past $ H_{u,l} $, we can learn a combination of different types of interest to reflect the user long-term interest (or state) before $ L_{u,l} $.

Instead of performing a summing operation to generate the query as in the original memory network~\cite{DBLP:conf/nips/SukhbaatarSWF15}, we apply a multi-dimensional attention model to generate the query embedding. This allows discriminating informative items that can better reflect the user preference to have a greater influence on the positioning of the corresponding external memory units. Formally, we denote the item embeddings in $ H_{u,l} $ as $ \mathbf{H}_{u,l} \in \mathbb{R}^{d \times |H_{u,l}|} $. The multi-dimensional attention to generate the query embedding $ \mathbf{z}_{u,l} $ is computed as:
\begin{equation}
\begin{aligned}
    \mathbf{H}_{u,l} &\coloneqq \mathbf{H}_{u,l} + \mathrm{PE}({H}_{u,l}) \,, \\
    \mathbf{S}_{u,l} = \mathrm{softmax}& \left(\mathbf{W}_{a}^{(3)} \tanh (\mathbf{W}_{a}^{(1)} \mathbf{H}_{u,l} +  (\mathbf{W}_{a}^{(2)} \mathbf{p}_{u}) \otimes \mathbf{1}_{\phi})\right) \,, \\    
    \mathbf{Z}_{u,l} &= \tanh(\mathbf{S}_{u,l} \cdot \mathbf{H}^{\top}_{u,l}) \,, \\
    \mathbf{z}_{u,l} &= \mathrm{avg}(\mathbf{Z}_{u,l}) \,,
\label{eq:multi_attention_query}
\end{aligned}
\end{equation}
where $ \mathrm{PE}(\cdot) $ is the sinusoidal positional encoding function that maps the item positions into position embeddings, which is the same as the one used in Transformer~\cite{DBLP:conf/nips/VaswaniSPUJGKP17}. 
$\phi$ equals to $|H_{u,l}|$,
$ \otimes $ denotes the outer product. $ \mathbf{W}_{a}^{(1)}, \mathbf{W}_{a}^{(2)} \in \mathbb{R}^{d \times d}$ and $ \mathbf{W}_{a}^{(3)} \in \mathbb{R}^{h \times d} $ are the learnable parameters in the attention model, and $ h $ is the hyper-parameter to control the number of dimensions in the attention model. $ \mathbf{S}_{u,l} \in \mathbb{R}^{h \times |H_{u,l}|} $ is the attention score matrix. $ \mathbf{Z}_{u,l} \in \mathbb{R}^{h \times d} $ is the matrix representation of the query, and each of the $h$ rows represents a different aspect of the query. Finally, $ \mathbf{z}_{u,l} \in \mathbb{R}^{d} $ is the combined query embedding that averages the different aspects.

Given the query embedding $ \mathbf{z}_{u,l} $, we use this query to find the appropriate combination of the shared user latent interest in the memory network. Formally, the keys and values of the memory network~\cite{DBLP:conf/nips/SukhbaatarSWF15,DBLP:conf/emnlp/MillerFDKBW16} are denoted as $ \mathbf{K} \in \mathbb{R}^{d \times m} $ and $ \mathbf{V} \in \mathbb{R}^{d \times m} $, respectively, where $ m $ is the number of memory units in the memory network. Therefore, the user long-term interest embedding can be modeled as:
\begin{equation}
\begin{aligned}
    s_i =& \; \mathrm{softmax}(\mathbf{z}^{\top}_{u,l} \cdot \mathbf{k}_i) \,, \\
    \mathbf{o}_{u,l} &= \sum_{i} s_i \, \mathbf{v}_i \,, \\
    \mathbf{p}^{H}_{u, l} &= \mathbf{z}_{u,l} + \mathbf{o}_{u,l} \,,
\label{eq:memory_bank}
\end{aligned}
\end{equation}
where $ \mathbf{k}_i, \mathbf{v}_i \in \mathbb{R}^{d} $ are the $ i $-th memory unit and the superscript $ H $ denotes the representation is from the user long-term interest.

\subsection{Interest Fusion}
We have obtained the user short-term interest representation and the long-term interest representation. The next aim is to combine these two kinds of hidden representations in the GNN framework to facilitate the user preference prediction on unrated items. Here, we modify Eq.~\ref{eq:GCN_aggregation2} to bridge the user short-term interest and long-term interest.

Specifically, we borrow the idea of LSTM~\cite{DBLP:journals/neco/HochreiterS97} that uses learnable gates to balance the current inputs and historical hidden states. Similarly, we propose a learnable gate to control how much the recent user interest representation and the long-term user interest representation can contribute to the combined user interest for item prediction:
\begin{equation}
\begin{aligned}
    & \mathbf{g}_{u,l} = \sigma\left(\mathbf{W}_{g}^{(1)} \cdot \frac{1}{|L|}\sum_{i \in L_{u,l}} \mathbf{h}_i + \mathbf{W}_{g}^{(2)} \cdot \mathbf{p}^{H}_{u,l} + \mathbf{W}_{g}^{(3)} \cdot \mathbf{p}_{u}\right) \,, \\
    & \mathbf{p}_{u,l}^{C} = \mathbf{g}_{u,l} \odot  \frac{1}{|L|}\sum_{i \in L_{u,l}} \mathbf{h}_i + (\mathbf{1}_d - \mathbf{g}_{u,l}) \odot  \mathbf{p}^{H}_{u,l} \,,
\label{eq:gate}
\end{aligned}
\end{equation}
where $ \mathbf{W}_{g}^{(1)}, \mathbf{W}_{g}^{(2)}, \mathbf{W}_{g}^{(3)} \in \mathbb{R}^{d \times d}$ are the learnable parameters in the gating layer, $ \odot $ denotes the element-wise multiplication, and $ \mathbf{g}_{u,l} \in \mathbb{R}^{d} $ is the learnable gate. The superscript $C$ denotes the fusion of long- and short-term interests.

\subsection{Item Co-occurrence Modeling}
Successful learning of pairwise item relationships is a key component of recommender systems due to its effectiveness and interpretability.  This has been studied and exploited in many recommendation models~\cite{DBLP:journals/tois/DeshpandeK04,DBLP:reference/sp/NingDK15}. In the sequential recommendation problem, the closely related items may appear one after another in the item sequence. For example, after purchasing a mobile phone, the user is much more likely to buy a mobile phone case or protector. To capture the item co-occurrence patterns, we use a bilinear function to explicitly model the pairwise relations between the items in $ L_{u,l} $ and other items. This function takes the form
\[
\mathbf{e}^{\top}_i \, \mathbf{W}_r \, \mathbf{q}_j \,,
\]
where $ \mathbf{W}_r \in \mathbb{R}^{d \times d} $  is a matrix of the learnable parameters that captures the correlations between item latent features.

\subsection{Prediction and Training}
To infer the user preference, we have a prediction layer to combine the aforementioned factors together:
\begin{equation}
    \hat{r}_{u,j} = \mathbf{p}_{u}^{\top} \cdot \mathbf{q}_{j} + \mathbf{p}^{C \top}_{u,l} \cdot \mathbf{q}_{j} + \frac{1}{|L|}\sum_{i \in L_{u,l}} \mathbf{e}^{\top}_i \, \mathbf{W}_r \, \mathbf{q}_j \,.
\end{equation}

As the training data is derived from the user implicit feedback, we optimize the proposed model with respect to the Bayesian Personalized Ranking objective~\cite{DBLP:conf/uai/RendleFGS09} via gradient descent. This involves optimizing the pairwise ranking between the positive (observed) and negative (non-observed) items:
\begin{equation}
\begin{aligned}
\operatorname*{arg\,min}_{\mathbf{U}, \mathbf{Q}, \mathbf{E}, \mathbf{\Theta}} \sum_{u} \sum_{l} & \sum_{(L_{u,l}, H_{u,l}, j, k)} -\log\sigma(\hat{r}_{u,j} - \hat{r}_{u,k}) +  \\ 
& \lambda(||\mathbf{P}||^{2} + ||\mathbf{Q}||^{2} + ||\mathbf{E}||^{2} + ||\mathbf{\Theta}||^{2}) \,.
\end{aligned}
\end{equation}
Here $ j $ denotes the positive item in $ T_{u,l} $, and $ k $ denotes the randomly sampled negative item, $ \sigma $ is the sigmoid function, $ \mathbf{\Theta} $ denotes other learnable parameters in the model, and $ \lambda $ is the regularization parameter. $ \mathbf{p}_{*} $, $ \mathbf{q}_{*} $ and $ \mathbf{e}_{*} $ are column vectors of $ \mathbf{P} $, $ \mathbf{Q} $ and $ \mathbf{E} $, respectively. When minimizing the objective function, the partial derivatives w.r.t.\ all parameters are computed by gradient descent with back-propagation.

\section{Evaluation}

In this section, we first describe the experimental set-up. We then report the results of conducted experiments and demonstrate the effectiveness of the proposed modules.

\subsection{Datasets}
The proposed model is evaluated on five real-world datasets from various domains with different sparsities: \textit{MovieLens-20M} \cite{DBLP:journals/tiis/HarperK16}, \textit{Amazon-Books} and \textit{Amazon-CDs} \cite{DBLP:conf/www/HeM16}, \textit{Goodreads-Children} and \textit{Goodreads-Comics} \cite{DBLP:conf/recsys/WanM18}. \textit{MovieLens-20M} is a user-movie dataset collected from the \textit{MovieLens} website; the dataset has 20 million user-movie interactions. The \textit{Amazon-Books} and \textit{Amazon-CDs} datasets are adopted from the Amazon review dataset with different categories, i.e., CDs and Books, which cover a large amount of user-item interaction data, e.g., user ratings and reviews. The \textit{Goodreads-Children} and \textit{Goodreads-Comics} datasets were collected in late 2017 from the \textit{goodreads} website with a focus on the genres of Children and Comics. In order to be consistent with the implicit feedback setting, we keep those with ratings no less than four (out of five) as positive feedback and treat all other ratings as missing entries on all datasets. To filter noisy data, we only keep the users with at least ten ratings and the items with at least ten ratings. The data statistics after preprocessing are shown in Table \ref{tab:data_statistics}. 

For each user, we use the earliest 70\% of the interactions in the user sequence as the training set and use the next 10\% of interactions as the validation set for hyper-parameter tuning. The remaining 20\% constitutes the test set for reporting model performance. Note that during the testing procedure, the input sequences include the interactions in both the training set and validation set. The learning of all the models is carried out five times to report the average results.

\begin{table}[ht]
\centering
\caption{\label{tab:data_statistics}The statistics of the datasets.}
\begin{tabular}{ |c|c|c|c|c|c| }
 \hline
 Dataset & \#Users & \#Items & \#Interactions & Density \\
 \hline
 \textit{CDs} & 17,052 & 35,118 & 472,265 & 0.079\% \\ 
 \hline
 \textit{Books} & 52,406 & 41,264 & 1,856,747 & 0.086\% \\ 
 \hline
 \textit{Children} & 48,296 & 32,871 & 2,784,423 & 0.175\% \\ 
 \hline
 \textit{Comics} & 34,445 & 33,121 & 2,411,314 & 0.211\% \\ 
 \hline
 \textit{ML20M} & 129,797 & 13,649 & 9,921,393 & 0.560\% \\ 
 \hline 
\end{tabular}
\end{table}

\begin{table*}[ht]
\centering
\caption{\label{tab:performance_comparison}The performance comparison of all methods in terms of \textit{Recall@10} and \textit{NDCG@10}. The best performing method is boldfaced. The underlined number is the second best performing method. * indicates the statistical significance for $ p <= 0.01 $ compared to the best baseline method based on the paired t-test.}
\begin{tabular}{|l|cc|cc|cc|cc|cc|}
\hline
\multirow{2}{*}{Methods} & \multicolumn{2}{c|}{CDs} & \multicolumn{2}{c|}{Books} & \multicolumn{2}{c|}{Children} & \multicolumn{2}{c|}{Comics} & \multicolumn{2}{c|}{ML20M} \\ \cline{2-11}
& R@10 & N@10 & R@10 & N@10  & R@10 & N@10 & R@10 & N@10 & R@10 & N@10  \\ \hline
BPRMF & 0.0269  & 0.0145 & 0.0260 & 0.0151 & 0.0814 & 0.0664 & 0.0788 & 0.0713  & 0.0774  & 0.0785  \\
\hline
GRU4Rec & 0.0302 & 0.0154 & 0.0266 & 0.0157 & 0.0857 & 0.0715 & 0.0958 & 0.0912 & 0.0804 & 0.0912 \\
GRU4Rec+ & 0.0356 & 0.0171 & 0.0301 & 0.0171 & 0.0978 & 0.0821 & 0.1288 & 0.1328 & 0.0904 & 0.0946  \\
GC-SAN & 0.0372 & \underline{0.0196} & 0.0344 & \underline{0.0256} & 0.1099 & 0.0967 & 0.1490 & 0.1563 & 0.1161 & \underline{0.1119} \\
\hline
Caser & 0.0297 & 0.0163 & 0.0297 & 0.0216 & 0.1060 & 0.0943 & 0.1473 & 0.1529 & \underline{0.1169} & 0.1116 \\
SARSRec & 0.0341 & 0.0193 & \underline{0.0358} & 0.0240 & \underline{0.1165} & \underline{0.1007} & \underline{0.1494} & \underline{0.1592} & 0.1069 & 0.1014 \\
MARank & \underline{0.0382} & 0.0151 & 0.0355 & 0.0223 & 0.1092 & 0.0980 & 0.1325 & 0.1431 & 0.1002 & 0.1031 \\
\hline
MA-GNN & \textbf{0.0442*} & \textbf{0.0214*} & \textbf{0.0432*} & \textbf{0.0279*} & \textbf{0.1215} & \textbf{0.1137*} & \textbf{0.1617*} & \textbf{0.1655} & \textbf{0.1236*} & \textbf{0.1272*} \\
Improvement & 15.7\% &  9.2\% & 20.7\% & 9.0\% & 4.3\% & 12.9\% &  8.23\% & 4.0\% & 5.7\% & 13.7\% \\
\hline 
\end{tabular}
\end{table*}

\subsection{Evaluation Metrics}
We evaluate all the methods in terms of \textit{Recall@K} and \textit{NDCG@K}. For each user, Recall@K (R@K) indicates what percentage of her rated items emerge in the top $ K $ recommended items. NDCG@K (N@K) is the normalized discounted cumulative gain at $ K $, which takes the position of correctly recommended items into account. $ K $ is set to $ 10 $.

\subsection{Methods Studied}
To demonstrate the effectiveness of our model, we compare to the following recommendation methods:
(1) \textbf{BPRMF}, Bayesian Personalized Ranking based Matrix Factorization \cite{DBLP:conf/uai/RendleFGS09}, a classic method for learning pairwise item rankings; 
(2) \textbf{GRU4Rec}, Gated Recurrent Unit for Recommendation \cite{DBLP:journals/corr/HidasiKBT15}, which uses recurrent neural networks to model item sequences for session-based recommendation;
(3) \textbf{GRU4Rec+}, an improved version of GRU4Rec \cite{DBLP:conf/cikm/HidasiK18}, which adopts an advanced loss function and sampling strategy;
(4) \textbf{GC-SAN}, Graph Contextualized Self-Attention Network \cite{DBLP:conf/ijcai/XuZLSXZFZ19}, which uses a graph neural network and self-attention mechanism for session-based recommendation;
(5) \textbf{Caser}, Convolutional Sequence Embedding Recommendation, \cite{DBLP:conf/wsdm/TangW18}, which captures high-order Markov chains via convolution operations;
(6) \textbf{SASRec}, Self-Attention based Sequential Recommendation \cite{DBLP:conf/icdm/KangM18}, which uses an attention mechanism to identify relevant items for prediction;
(7) \textbf{MARank}, Multi-order Attentive Ranking model~\cite{DBLP:conf/aaai/YuZLZ19}, which unifies individual- and union-level item interactions to infer user preference from multiple views;
(8) \textbf{MA-GNN}, the proposed model, which applies a memory augmented GNN to combine the recent and historical user interests and adopts a bilinear function to explicitly capture the item-item relations.

\subsection{Experiment Settings}
In the experiments, the latent dimension of all the models is set to 50. For the session-based methods, we treat the items in a short-term window as one session. For GRU4Rec and GRU4Rec+, we find that a learning rate of $ 0.001 $ and batch size of $ 50 $ can achieve good performance. These two methods adopt Top1 loss and BPR-max loss, respectively. For GC-SAN, we set the weight factor $ \omega $ to $ 0.5 $ and the number of self-attention blocks $ k $ to $ 4 $. For Caser, we follow the settings in the author-provided code to set $ |L|=5 $, $ |T|=3 $, the number of horizontal filters to $ 16 $, and the number of vertical filters to $ 4 $. For SASRec, we set the number of self-attention blocks to $ 2 $, the batch size to $ 128 $, and the maximum sequence length to $ 50 $. For MARank, we follow the original paper to set the number of depending items as $ 6 $ and the number of hidden layers as $ 4 $. The network architectures of the above methods are configured to be the same as described in the original papers. The hyper-parameters are tuned on the validation set.

For MA-GNN, we follow the same setting in Caser to set $ |L|=5 $ and $ |T|=3 $. Hyper-parameters are tuned by grid search on the validation set. The embedding size $ d $ is also set to $ 50 $. The value of $ h $ and $ m $ are selected from $ \{5, 10, 15, 20 \} $. The learning rate and $ \lambda $ are set to $ 0.001 $ and $ 0.001 $, respectively. The batch size is set to $ 4096 $.

\subsection{Performance Comparison}
The performance comparison results are shown in Table \ref{tab:performance_comparison}. 

\textbf{Observations about our model}. First, the proposed model, MA-GNN, achieves the best performance on all five datasets with all evaluation metrics, which illustrates the superiority of our model. Second, MA-GNN outperforms SASRec. Although SASRec adopts the attention model to distinguish the items users have accessed, it neglects the common item co-occurrence patterns between two closely related items, which is captured by our bilinear function. Third, MA-GNN achieves better performance than Caser, GC-SAN and MARank. One major reason is that these three methods only model the user interests in a short-term window or session, but fail to capture the long-term item dependencies. On the contrary, we have a memory network to generate the long-term user interest. Fourth, MA-GNN obtains better results than GRU4Rec and GRU4Rec+. One possible reason is that GRU4Rec and GRU4Rec+ are session-based methods that do not explicitly model the user general interests. Fifth, MA-GNN outperforms BPRMF. BPRMF only captures the user general interests, and does not incorporate the sequential patterns of user-item interactions. As such, BPRMF fails to capture the user short-term interests.

\textbf{Other observations}. First, all the results reported on MovieLens-20M, GoodReads-Children and GoodReads-Comics are better than the results on other datasets. The major reason is that the other datasets are sparser and data sparsity negatively impacts recommendation performance. Second, MARank, SASRec and GC-SAN outperform Caser on most of the datasets. The main reason is that these methods can adaptively measure the importance of different items in the item sequence, which may lead to more personalized user representation learning. Third, Caser achieves better performance than GRU4Rec and GRU4Rec+ in most cases. One possible reason is that Caser explicitly inputs the user embeddings into its prediction layer, which allows it to learn general user interests. Fourth, GRU4Rec+ performs better than GRU4Rec on all datasets. The reason is that GRU4Rec+ not only captures the sequential patterns in the user-item sequence but also has a superior objective function---\textit{BPR-max}. Fifth, all the methods perform better than BPR. This illustrates that a technique that can only perform effective modeling of the general user interests is incapable of adequately capturing the user sequential behavior.

\begin{table}[ht]
\centering
\caption{\label{tab:ablation_analysis}The ablation analysis. \textit{S} denotes the short-term interest module, \textit{H} denotes the long-term interest module, \textit{concat} denotes the concatenation operation.}
\begin{tabular}{ |l|c|c|c|c| }
\hline
\multirow{2}{*}{Architecture} & \multicolumn{2}{c|}{\textit{CDs}} & \multicolumn{2}{c|}{\textit{Books}} \bigstrut \\\cline{2-5} 
& R@10 & N@10 & R@10 & N@10 \bigstrut \\ 
\hline
(1) MF & 0.0269 & 0.0145 & 0.0310 & 0.0177 \\
(2) MF+S & 0.0306 & 0.0158 & 0.0324 & 0.0185 \\
(3) MF+S+H+gating & 0.0401 & 0.0191 & 0.0351 & 0.0208 \\
(4) MF+S+H+concat & 0.0295 & 0.0148 & 0.0296 & 0.0206 \\
(5) MF+S+H+GRU & 0.0268 & 0.0147 & 0.0306 & 0.0204 \\
(6) MA-GNN & \textbf{0.0442} & \textbf{0.0214} & \textbf{0.0432} & \textbf{0.0279} \\
\hline
\end{tabular}
\end{table}

\subsection{Ablation Analysis} \label{sec:ablation}
To verify the effectiveness of the proposed short-term interest modeling module, long-term interest modeling module, and item co-occurrence modeling module, we conduct an ablation study in Table \ref{tab:ablation_analysis}. This demonstrates the contribution of each module to the MA-GNN model. In (1), we utilize only the BPR matrix factorization without other components to show the performance of modeling user general interests. In (2), we incorporate the user short-term interest by the vanilla graph neural network (Eq.~\ref{eq:GCN_aggregation1} and~\ref{eq:GCN_aggregation2}) on  top of (1). In (3), we integrate the user long-term interest with the short-term interest via the proposed interest fusion module (Eq.~\ref{eq:multi_attention_query},~\ref{eq:memory_bank} and~\ref{eq:gate}) on top of (2). In (4), we replace the interest fusion module in (3) with the concatenation operation to link the short-term interest and long-term interest. In (5), we replace the concatenation operation with a gated recurrent unit~\cite{DBLP:conf/emnlp/ChoMGBBSB14} (GRU). In (6), we present the overall MA-GNN model to show the effectiveness of the item co-occurrence modeling module.

From the results shown in Table \ref{tab:ablation_analysis}, we make the following observations. First, comparing (1) and (2)-(6), we can observe that although the conventional BPR matrix factorization can capture the general user interests, it cannot effectively model the short-term user interests. Second, from (1) and (2), we observe that incorporating the short-term interest using the conventional aggregation function of the GNN slightly improves the model performance. Third, in (3), (4) and (5), we compare three ways to bridge the user short-term interest and long-term interest. From the results, we can observe that our proposed gating mechanism achieves considerably better performance than concatenation or the GRU, which demonstrates that our gating mechanism can adaptively combine these two kinds of hidden representations. Fourth, from (3) and (6), we observe that by incorporating the item co-occurrence pattern, the performance further improves. The results show the effectiveness of explicitly modeling the co-occurrence patterns of the items that a user has accessed and those items that the user may interact with in the future. The item co-occurrence pattern can provide a significant amount of supplementary information to help capture the user sequential dynamics.

\subsection{Influence of Hyper-parameters}
\begin{figure}[t!]
    \centering
    \begin{subfigure}[t]{0.25\textwidth}
        \centering
        \includegraphics[width=\linewidth]{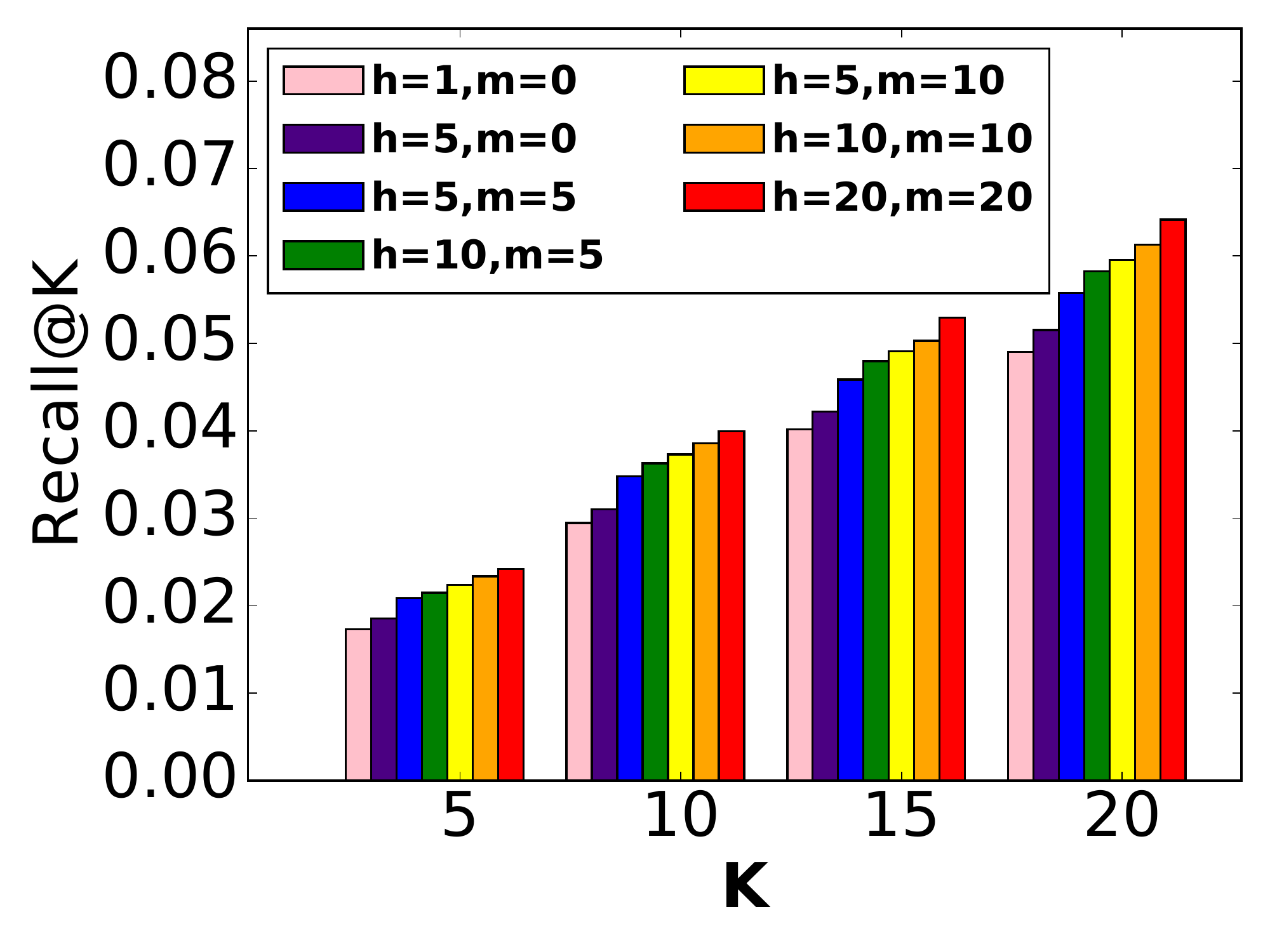}
        \caption{\label{fig:Books_dim_var}$ h $ and $ m $ on CDs}
    \end{subfigure}%
    \begin{subfigure}[t]{0.25\textwidth}
        \centering
        \includegraphics[width=\linewidth]{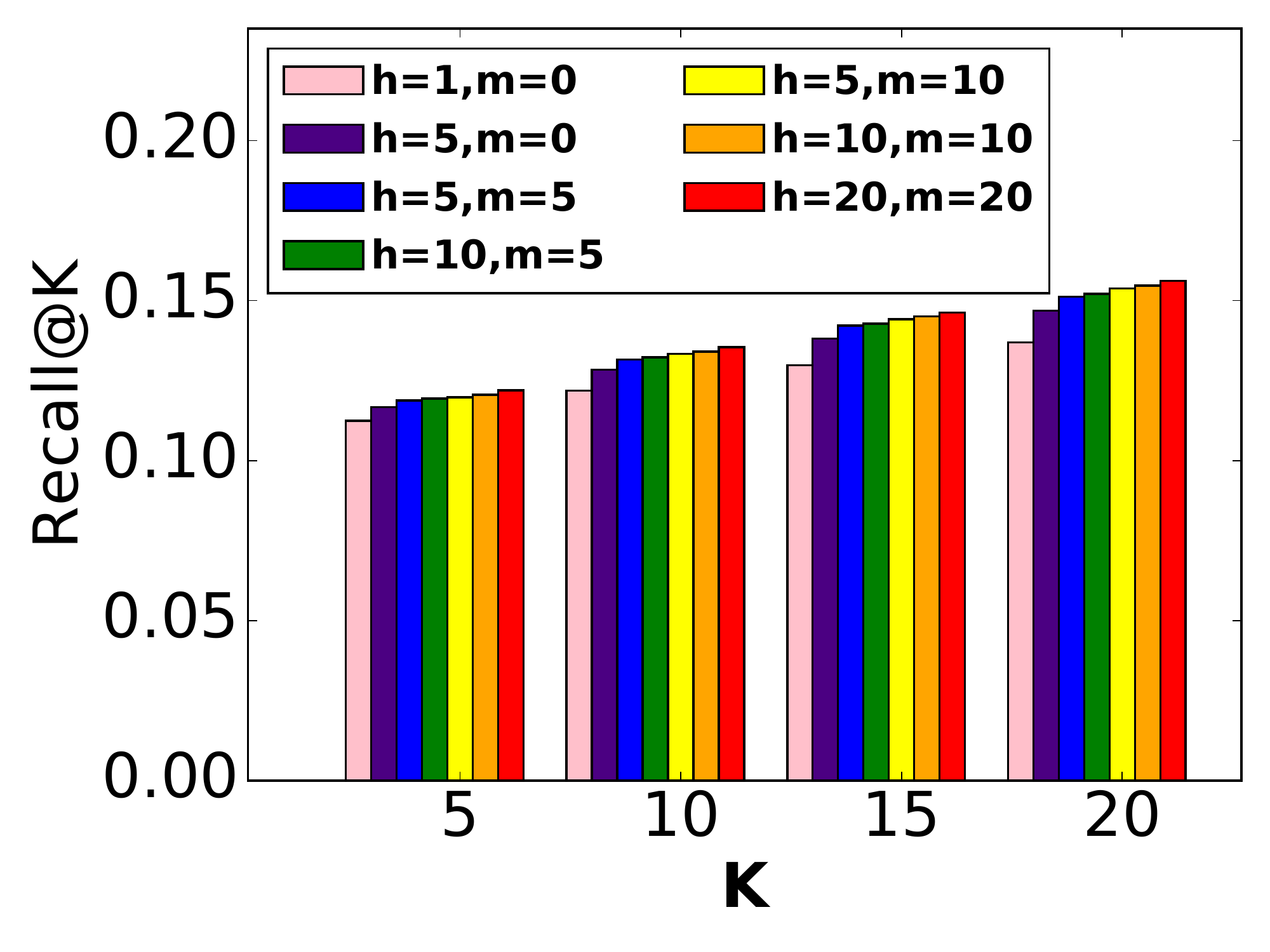}
        \caption{\label{fig:CDs_dim_var}$ h $ and $ m $ on Comics}
    \end{subfigure}
    \caption{\label{fig:hyper_parameter}The variation of $ h $ and $ m $.}
\end{figure}

The dimension $ h $ of the multi-dimensional attention model and the number $ m $ of the memory units are two important hyper-parameters in the proposed model. We investigate their effects on CDs and Comics datasets in Figure~\ref{fig:hyper_parameter}.

From the results in Figure~\ref{fig:hyper_parameter}, we observe that both the multi-dimensional attention and the memory network contribute to capturing the long-term user interests. These two components lead to a larger improvement in performance for the CDs dataset compared to the Comics dataset, indicating that they may help to alleviate the data sparsity problem.

\begin{figure}[ht]
    \centering
    \includegraphics[width=0.5\textwidth]{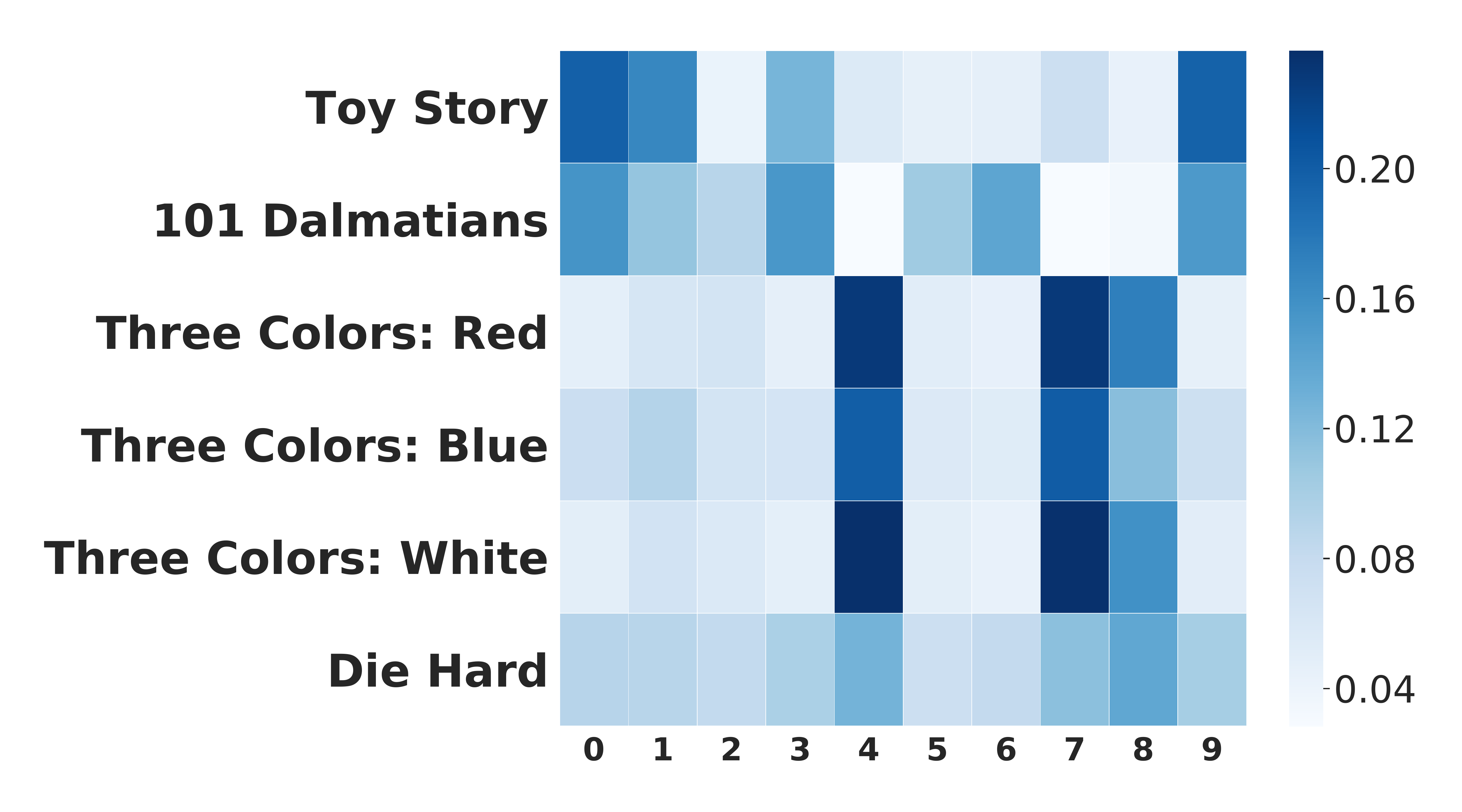}
    \caption{The attention visualization of memory networks.}
    \label{fig:heat_map}
\end{figure}

\subsection{Memory Visualization}
To validate whether each memory unit can represent a certain type of user interests, we conduct a case study on the MovieLens dataset to verify how each memory unit functions given different movies. We randomly select a user and several movies she watched. For simplicity, we treat each selected item as a query to visualize the attention weight $ s_i $ computed by Eq.~\ref{eq:memory_bank}. In this case, we set the number of memory units $ m $ to $ 10 $.

From Figure~\ref{fig:heat_map}, we observe that our memory units perform differently given different types of movies, which may illustrate that each memory unit in the memory network can represent one type of the user interest. For example, the \textit{Three Colors} trilogy has quite similar attention weights in the memory network, since these three movies are loosely based on three political ideals in the motto of the French Republic. \textit{Die Hard} is an action thriller movie, which is distinct from any other movies in the case study, explaining why it has a different weight pattern.

\section{Conclusion}
In this paper, we propose a memory augmented graph neural network (MA-GNN) for sequential recommendation. MA-GNN applies a GNN to model items' short-term contextual information, and utilize a memory network to capture the long-range item dependency. In addition to the user interest modeling, we employ a bilinear function to model the feature correlations between items. Experimental results on five real-world datasets clearly validate the performance advantages of our model over many state-of-the-art methods and demonstrate the effectiveness of the proposed modules.

\bibliographystyle{aaai}
\bibliography{references.bib}

\end{document}